\documentclass{elsart}
\usepackage{graphicx}
\journal{Nuclear Instruments and Methods ...}
\begin{document}
\begin{frontmatter}
\title{Elements Discrimination in the Study of Super-Heavy Elements
using an Ionization Chamber.}
\author[jag,lpc]{A. Wieloch \corauthref{cor}}, \corauth[cor]{Corresponding author}
\ead{ufwieloc@cyf-kr.edu.pl}
\author[jag,lpc]{Z. Sosin},
\author[lpc]{J. P\'e{}ter},
\author[jag]{K. \L{}ojek},
\author[sac]{N. Alamanos},
\author[lpc]{N. Amar},
\author[gan]{R. Anne},
\author[lpc]{J.C. Ang\'e{}lique},
\author[gan]{G. Auger},
\author[sac]{R. Dayras}, 
\author[sac]{A. Drouart}, 
\author[lpc]{J.M. Fontbonne}, 
\author[sac]{A. Gillibert},
\author[lpc]{S. Gr\'e{}vy},
\author[ulb]{F. Hanappe}, 
\author[cen]{F. Hannachi},
\author[gan]{R. Hue},
\author[gan]{A. Khouaja },
\author[lpc]{T. Legou},
\author[csn]{A. Lopez-Martens},
\author[lpc]{E. Li\'e{}nard}, 
\author[lpc]{L. Manduci},
\author[gan]{F. de Oliveira Santos},
\author[cat]{G. Politi}, 
\author[gan]{M.G. Saint-Laurent},
\author[gan]{C. Stodel}, 
\author[str]{L. Stuttg\'e{}},
\author[lpc]{J. Tillier},
\author[gan]{R. de Tourreil},
\author[gan]{A.C.C. Villari},
\author[gan]{J.P. Wieleczko}
\author[]{(FULIS Collaboration)}
\address[jag]{M. Smoluchowski Institute of Physics, Jagellonian University,
 Reymonta 4, 30-059 Krak\'o{}w, Poland}
\address[lpc]{LPC-ENSI Caen, Bld Mar\'e{}chal Juin, 14050 Caen Cedex, France}
\address[sac]{CEA-Saclay, DAPNIA-SPhN, 91191 Gif sur Yvette Cedex, France}
\address[gan]{GANIL, B.P. 5027, 14076 Caen Cedex 5, France}
\address[ulb]{PNTPM, Universit\'e{} Libre de Bruxelles, Belgique}
\address[cen]{ENBG, BP 120,33175 Gradignan, France}
\address[csn]{CSNSM, Orsay, France}
\address[cat]{Istituto Fisica dell'Universit\`a{}, Catania, Italy}
\address[str]{IreS, Strasbourg, France}

\begin{abstract}
   Dedicated ionization chamber was built and installed to measure the energy 
loss of very heavy nuclei at 2.7 MeV/u produced in fusion reactions in inverse 
kinematics (beam of $^{208}$Pb).
After going through the ionization chamber, products of reactions on  
$^{12}$C, $^{18}$O targets are implanted in a Si detector. Their identification 
through their alpha decay chain is ambiguous when their half-life is short. 
After calibration with Pb and Th nuclei, the ionization chamber signal allowed 
us to resolve these ambiguities. In the search for rare super-heavy nuclei 
produced in fusion reactions in inverse or symmetric kinematics, such a 
chamber will provide  direct information on the nuclear charge of each 
implanted nucleus. 
\end{abstract}

\end{frontmatter}

\section {Introduction.}

 	The production of  super heavy elements (SHE)  and the study of their properties are 
conducted in  several research centers (ANL, GSI Darmstadt, JINR  Dubna, LBL Berkeley, 
GANIL Caen,  Riken, Jyv\"a{}skyl\"a{}). So far all experiments were done in normal 
kinematics (projectile much lighter than the target nucleus), which means that the 
recoiling fusion nuclei and their detected evaporation residues have very low kinetic 
energies per nucleon, typically in the 0.2-0.5 MeV/u range. Most of the groups use very 
similar detection and identification techniques. A typical detection setup consists of 2 or 3 
carbon foils associated with micro-channel plate detectors which give a time-of-flight 
information (i.e. the velocity ), and a position silicon detector where the  reaction 
products are implanted  and which returns information on the energy and x-y position. Due 
to the very large mass of the evaporation residues, the ionization defect in the Si 
implantation detector is very large,  the energy signal is small and has a broad 
distribution (due to fluctuations of the energy loss in the carbon foils 
and the Si crystal response), so that mass 
determination through kinetic energy and velocity is not feasible. Neither it is 
possible to install a $\Delta$E detector (thin Si detector or gas chamber) since it would stop 
the evaporation residues. Therefore  no direct identification of the detected reaction 
product is made. Actually, the fusion evaporation residues and other reaction 
products are identified via alpha-particle decay chains leading to a known nucleus 
\cite{Hofm:95:1}-\cite{Pete:01:1}. Indirect although unambiguous identification 
is done. The observed  $\alpha$-chain or fission decay has to be correlated with
the signal from 
the heavy ion in the implantation detector. This technique is powerful but 
suffers from two limitations: i) 
the detection efficiency for $\alpha$-particles is not 100 \% , so the chains are often not 
complete. ii) since the evaporation residue cross section becomes very small for 
compound nuclei above Z$\sim$106, there are very few such products (1-2 per day or per 
week) among a background due to other reaction products (transfer reactions) some 
of which are $\alpha$-emitters, so the probability of wrong chain reconstruction and  
identification may be large \cite{Zlo:00:1}, \cite{Sch:00:1}. Therefore any direct information on 
the mass and/or charge of the implanted product would be very useful, even though it 
does not offer a resolution of 1 mass or charge unit. In the detection of rare events, it 
is clear that redundant information is needed in order to eliminate erroneous 
identifications. Each additional independent physical quantity that can be measured 
and delivered by a modified detection system is precious. 

Such information can be obtained in the case of inverse kinematics or nearly 
symmetric systems. Indeed the compound nucleus recoil velocity is much larger. It 
was calculated that an estimate of the mass can be obtained after calibration of the 
implantation detector response with heavy projectiles at the compound nucleus 
velocity and with known very heavy nuclei or ``light'' super-heavy fusion nuclei (formed 
with a  relatively large  cross section). Similarly, a thin detector (gas counter) can be  installed
before the implantation detector to provide a $\Delta$E  information, so that an estimate of  
the nuclear charge  can be obtained. In GANIL a test experiment was performed with 
such a detection set-up in inverse kinematics: beam of $^{208}$Pb at 5 MeV/u on light 
targets.

In this paper we show that the $\Delta$E signal from an ionization chamber can be 
used to distinguish between different reaction channels leading to SHE. In  chapter 
two we present the detection setup of the GANIL experiment. Next we describe the 
construction and properties of the ionization chamber. In chapter four some results 
from the reactions $^{208}$Pb+$^{12}$C and $^{208}$Pb+$^{18}$O  are presented stressing 
the role of the signals received from the $\Delta$E detector. The last section  gives 
conclusions and some perspectives.

\section {Experimental detection setup.} 
\label {detsetup}
The main tool of SHE research at GANIL is the Wien filter at the LISE magnetic
spectrometer with crossed magnetic and 
electric fields (figure \ref{fig1}). 
It is divided into two identical halves. The ratio of magnetic to electric field is set 
to select the velocity of the complete fusion fragments (evaporation residues). 
Their values are chosen 
so that the projectiles are deflected in the first half of the Wien filter and stopped in a Faraday plate. 
Two pairs of independently movable slits and a beam profiler were installed at mid-
filter. The suppression of unwanted products is improved by a dipole magnet  located 
after the velocity filter.
	
The reaction chamber contains several rotating targets: a 670 mm diameter wheel 
bearing 35 targets and rotating at 2000 RPM, allowing targets with a low 
melting temperature to sustain intense beams, a 
16 cm wheel with 8 targets for materials having a higher melting temperature and small 
rotating targets (diameter 5 cm) used for calibration at low intensity. The time 
structure of the beam is synchronized with the rotation. A Si detector continuously 
monitors the status of each target. 
To equilibrate the charge state of the fusion products before they enter the Wien filter, 
a 50 $\mu$m carbon stripper was placed 40 cm behind the target. 
After the dipole magnet, the velocity of each product is obtained 
from the ToF between 2 aluminized mylar foils and micro-
channel plate detectors. Then an ionization chamber provides a $\Delta$E signal. It is 
described in the next section. The residual kinetic energy and localization are given 
by a  X-Y Si implantation detector (IMP).

The energy of  $\alpha$-particles and fission fragments escaping from IMP is measured 
with a set of 8 Si detectors which form a tunnel. A veto Si detector  is installed behind IMP to reject 
light particles which might punch through it. Special electronics and data acquisition 
systems were developed. A fast analysis program allowed us to identify $\alpha$-decay 
chains or spontaneous fission events on line.

The transmission of fusion evaporation residue produced at the target position through the 
quadrupoles, filter and final dipole was studied via a simulation code. The optics 
was checked and the Wien filter was calibrated using $^{208}$Pb ions at several energies, 
which were also used for calibrating the ToF and IMP response. The 
transmission efficiency of the whole set-up was checked via fusion reactions with 
known cross sections and  $\alpha$-decay chains.

Direct identification of the SHE is done in the two-dimensional plot:
energy signal from the IMP vs ToF signal. In the ToF range selected by 
the velocity filter, there is always a 
background of scattered projectiles, $^{208}$Pb ions in this case. 
The higher energy signals come from heavier
products. This informs us on the mass number of the reaction product. The energy 
and ToF resolutions are such that the mass resolution is  several  \%. This is sufficient to 
eliminate  many of the heavy transfer products. An additional discrimination 
between products with different Z's is  necessary to improve the direct product
identification.

\section {Construction and specification of the ionization chamber (IC).}

For the production of super-heavy nuclei via complete fusion followed by evaporation of 1
 or a few neutrons (detected evaporation residues, ER), projectiles at (about) 5 
MeV/u are used.  In inverse kinematics experiments  (Pb, U projectiles) 
the kinetic energy of the ER's can be as high as 2.5 - 3 MeV/u. In nearly 
symmetric systems (Xe + Sn), it is around 1.2 MeV/u. Such energy is 
sufficient to install a  $\Delta$E detector; here it 
is an ionization chamber. 

The process of how an ion with charge Z loses energy passing through matter is 
quite complicated and has been a subject which received great theoretical and 
experimental interest. In the few MeV/u energy region mainly electron interactions are 
responsible for the energy losses while the nuclear interactions are negligible. Here 
the effective charge becomes the important quantity. The Bethe-Bloch formula is no 
longer exact and is usually replaced by semi-empirical ones \cite{Zieg:85:1,Hube:89:1}. 
Experimental data shows that in the vicinity of the Bragg maximum, energy loss is 
approximately proportional to Z. This dependence can be used to identify  the charge 
of the ion. For our 
experiment we are interested in discriminating between nuclei with charge Z=82 
(projectiles) to 88 (fusion products on $^{12}$C material used to support the target 
material and as stripper foils)  and super-heavy compound nuclei (Z$\geq$106). A  resolution in 
Z of 1 \% would allow us to measure Z of the ER.  We do not expect such a resolution, but, as one 
can see, such a detector will be useful if its $\Delta$E (charge) resolution   is only  on the 
level of 10\% or better. 

The main factors which influence resolution of the gas detector are:
\begin{enumerate}
\item 
Detector thickness- $\Delta$x.  The energy loss $\Delta$E=$\Delta$x$\cdot$dE/dx  is proportional to the 
detector thickness, and straggling of the $\Delta$E is proportional to 
$\Delta x^{1/2}$ 
($\delta (\Delta E)\propto \Delta x^{1/2}$) \cite{england}. That means the 
relative resolution ($\delta (\Delta E)/ \Delta E $) is 
inversely proportional to $\Delta x^{1/2}$. 
\item
Non-uniformity of the detector thickness. In the case of the gas detector non 
uniformity is due to the window distortions caused by the gas pressure. This 
effect can by partially corrected by installing additional field wires which 
define a constant useful length (see below).
\item
Non-completeness of charge collection. Part of the charge generated by the 
passing particle is not collected by the detector mainly due to the 
recombination process. This effect decreases the detector resolution. If charge 
collection conditions determined by the electric field are different for various 
ion trajectories, the recombination process worsens the detector resolution 
even more.   
\item
Noise of the associated electronics.  The detector signal is broadened by the 
electronics noise. This broadening depends on the internal noise of the 
preamplifier and detector capacity. 
\end{enumerate}

Considering the above factors we decided to build an ionization chamber as a $\Delta$E 
detector.  In such a solution the amount of collected charge is almost 
insensitive to the anode voltage. This property is a result of absence of avalanche 
processes, which would be responsible for additional fluctuations of the collected 
charge.  The detector construction is schematically depicted in fig.\ref{fig1}. The electric field in 
the counter is generated by 3 planes of  wire grids. The distance between
consecutive planes is 1.5 cm and each one has the same 
dimensions: 5 cm x 5 cm. Within a grid wires are spaced 2 mm apart and their diameter 
is 10 $\mu$m. Such a solution minimizes the electric capacity of the counter as a 
consequence, electronic noise is reduced \cite{Sosi:94:1}. The 2 outer grids are
grounded while the middle one is set to a positive voltage around 150 V. 
The electric field in the counter is parallel to the trajectory of an entering ion. 
Because of this, the collection of the generated charge is independent on the point of the ion 
injection. Furthermore, for any given particle trajectory in the IC, the effect of
the generated charge on the electric field is identical. This means 
the same conditions are obtained for the collection of the useful  charge. As a result, 
the rise-time  and the shape of  the pulse  are very similar for each detected particle. 
It is therefore possible to optimize the shaping times in  a spectroscopic amplifier. 
Additionally, in order to 
minimize electronics noises, the low noise charge preamplifier was connected 
directly to the counter output signal \cite{Loje:03:1}. The charge
sensitivity of the preamplifier is 940 mV/pC. 

The counter thickness is determined by the distance between mylar windows and by 
the type and the pressure of gas. As said above the gas pressure causes a smearing of 
the counter thickness. It becomes dependent on the particle trajectory. The cathode grids 
put just besides the windows minimize this effect. If the window and the 
screening grid are at the same potential (ground), the electric field between 
them is approximately equal to zero. As a result, the charge generated in the region 
of non constant thickness (window-grid to cathode), do not influence the measured 
pulses. The charge from such a region recombines or is collected by the residual 
field. If the residual field collects this charge the collection time is long and a slowly 
increasing component appears in output pulse, but this slow component is eliminated 
by the spectroscopy amplifier. 
The effective transmission for the system due to the window support wires and three planes of wires 
grids was calculated to be around 96\%. 

This ionization chamber is operating with circulating isobutane under the 
pressure of 30 mbar. Circulation of the gas is controlled by a GANIL gas flow control 
system.  Mylar windows of 2.5 $\mu$m on both sides are used.

\section {Reactions Pb+$^{12}$C,$^{18}$O. Alpha chains and $\Delta$E signals.}

The resolution of the IC is  poorer than the difference in the energy 
loss of adjacent elements passing through the chamber. In order to 
study this resolution one has to utilize independent information on the
Z of the detected ion. For this purpose we use measurements
performed to calibrate the detection set-up. These measurements were done
with a 5 MeV/u $^{208}$Pb beam, delivered by the 1st stage cyclotron 
of the GANIL facility, impinging on $^{18}$O and  $^{12}$C  targets.

 With $^{18}$O target, the compound nucleus is $^{226}$Th$^*$ and has excitation
 energy E$^*$=35 MeV. Reaction products give  
$\Delta$E signals in the IC and their $\alpha$ radioactivity 
is used to identify  them after  they implant in the position-sensitive Si 
detector IMP(see sec.\ref{detsetup}).  In fig.\ref{fig2}  is shown the 
2-dimensional energy spectrum of $\alpha$ particles which are emitted at the same 
position as an  ER implanted in the strip detector. The figure shows that
there are well defined groups of $\alpha$'s. Using a standard ER-$\alpha$-$\alpha$ 
correlation technique alpha chains which start from the parent isotopes 
$^{221}$Th and $^{223}$Th can be identified. This means that the compound 
nucleus $^{226}$Th$^*$ which de-excites by 5n and 3n neutron evaporation channels  was  
produced. In the figure we marked also  
alphas that are emitted by daughter nuclei.  The parent nucleus 
(implanted residue) is unambiguously identified here because  the half life 
of the first emitted $\alpha$ 
from the residue is  long enough and these alphas can be detected by the 
detection system. The 4n channel, $^{222}$Th, must also be present, but    
the half-life of its daughter 
$^{218}$Ra$^*$  is 25 $\mu$s,  below the time resolution, 50$\mu$s, 
of the detection system; for the grand-daughter $^{214}$Rn$^*$, it is less than 1 $\mu$s.  
Conversely, the grand-grand-daughter  $^{210}$Po$^*$  has a much too long 
half-life: 138 days. Therefore  we cannot see  them. Only one alpha from the 
radioactive decay chain of $^{222}$Th  was identified. This alpha is emitted by 
the  $^{222}$Th itself. It has energy 7.98, 7.60 MeV and half live 2.5 ms which 
lies in the most convenient range of ms-0.1 s acceptable by our detection set up.  
Background pulses in the IC originate mainly from  scattered Pb ions with velocities matching 
the velocity  filter acceptance. 
This background may broaden the $\Delta$E distribution of other detected Z's. Indeed, if one  
Pb ion  implants in the IMP detector at the same x-y position as an ER before 
this ER emits its alpha, then the $\Delta$E signal of the Pb ion is attributed to the ER. 
This contribution depends on the counting rate. Because we use $\alpha$ chains 
to identify Z of the ER, it also depends on the time after which the first $\alpha$ is 
emitted. Such contribution  from the Pb background is given approximately by the 
formula r/($\lambda$+r), where r denotes the counting rate at a given x, y position on 
the IMP detector and $\lambda$ is a decay constant that determines life time of the ER. 
Taking this into account we estimate the background contribution for the Z=90 case on 
the level of few \%.

In fig.\ref{fig3} we present measured distributions
of the energy loss for different nuclei. The empty squares are for Z=90, 
while full circles are for Z=82. Of course, the distribution for Z=82 (Pb) ions comes from 
the analysis of the Pb ions which pass the filter. Solid lines in the figure are gaussian fits. 
The FWHM for both cases is about 6\%. 
Mean values extracted from two $\Delta$E distributions (Z=82, 90) are 
presented in the upper panel. The straight line is drawn to interpolate
the region of Z's between these two cases and also to extrapolate to higher elements. 
The energy loss is a function of the ion velocity but as it is shown on the 
fig.\ref{fig4} this dependence on the ion velocity is very weak in the considered energy region. 
So this line can be used as a ``calibration'' line.
 
For reactions on the carbon target, the compound nucleus is $^{220}$Ra$^*$
(E$^*$=23 MeV)  and we tried 
to select $\alpha$ 
chains to identify the ER's,  like for oxygen. But in this case the interpretation of the data is more 
complicated. In the expected $\alpha$ chains following xn de-excitation after 
the compound nucleus formation, the 1n chain starting from $^{219}$Ra 
(half-life: 10ms) and ending at $^{211}$Po is easily identified. In addition, 
there are alphas with very short emission time, especially the first one in the 
2n, 3n and 4n chains: $^{218}$Ra: 25.6 $\mu$s, $^{217}$Ra:  1.6 $\mu$s, $^{216}$Ra: 2/180 ns. 
If these chains  are populated  the alpha energy piles up with the ER kinetic energy pulse
and they are missed by the IMP detector. Because of that we can only identify 
the emitters of the observed alphas and cannot identify the implanted ion. 
The incident energy we used was chosen for other targets and is a few MeV below the 
Bass barrier for fusion with $^{12}$C. Measurements of 3, 4 and 5 n channels  
were made at higher incident energies \cite{Nomu:73:1}. Extrapolation to our 
energy (4.85 MeV/u at mid-target) leads to expect less than a few mb in the 
2n channel, i.e. $^{218}$Rn which we cannot identify. 1n and 3n channels could be 
present with much smaller cross sections. 1n ($^{219}$Ra) could be identified 
but no event was found.  The excitation energy is too low for feeding the 4n 
channel; anyway $^{215}$Ra has a quite convenient half-life (1.6 ms)  but 
unfortunately its daughter
$^{211}$Rn has a too long half-life: 14.6 days, so  no chain could be seen. 
During the analysis the only events we  could unambiguously identify 
were the alpha-chains going through $^{211}$Po or $^{213}$Rn. 
In the case of  $^{211}$Po  there are two possible candidates as implanted ER's : $^{215}$Rn - 
residue from the two $\alpha$ transfer reaction followed by evaporation of 1 neutron, or 
$^{211}$Po itself - residue from 
one $\alpha$ transfer followed by evaporation of 1 neutron . For the $^{213}$Rn (25 ms)
emitter the implanted father can be  $^{217}$Ra (1.6 $\mu$s) i.e.  3n channel  after 
fusion or  just $^{213}$Rn - from two $\alpha$ transfer reaction. One can 
not affirm which scenario is the right one. For Rn the background contribution is
negligible while for Po it is of the order of 20\%.

In order to determine what was
really implanted in these  two cases we plot the energy loss distribution as shown
in fig.\ref{fig3}:  open circles for implants associated with $^{211}$Po and full
squares for implants which produce the $^{213}$Rn emitter. Mean values from these two
distributions were then marked on the ``calibration'' line (upper  
picture in the same figure) assuming that  open 
circles and full squares correspond to Z=84 and Z=86 respectively.   These two points fit well on the line. 
In this way we may settle that the implanted nuclei are  exactly the 
emitters which were identified from alpha-residue correlations. We conclude that, as expected,  
the fusion process has  a low cross section  at this energy. 1 or 2  alpha particle transfer 
reactions are favored (3 alpha structure of $^{12}$C).
  
In figure \ref{fig3} we have put also the Gaussian line which corresponds
to the nucleus Z=114. As one can see from the predicted location of 
the line the IC should be able to  separate Z=114 from most actinide products.  

\section {Conclusions and remarks.}

 The specific thin ionization chamber, we developed,  was found to be very 
useful and quite powerful for resolving ambiguities in the identification of 
some nuclei produced in reactions of  $^{208}$Pb projectiles on a $^{12}$C 
target.
	
	This detector will be used in experiments on the production of 
super-heavy nuclei via fusion reactions in inverse or symmetric kinematics. 
It will provide a signal related to the nuclear charge of the detected nuclei. 
Although element resolution is not achieved, this signal will make it possible 
to eliminate most of the "background" of other reaction products. Also in 
inverse or symmetric kinematics, a large kinetic energy signal (several 
hundreds of MeV) will be obtained in the implantation detector.  After 
calibration of this signal and of the time-of-flight with heavy projectiles and 
evaporation residues with known velocities, a rough value of the mass will  
be obtained. Together, these values of mass and charge will provide direct and 
immediate information on the implanted nucleus, independently of its 
subsequent decay, thus reducing very much the number of erroneous 
identifications and strongly improving  the quality of the data.

Acknowledgments: The authors want to thank all the technicians and 
engineers who made it possible to overcome the specific difficulties 
of these experiments: the electronics, detectors and mechanics groups 
at LPC Caen, Ganil and Dapnia Saclay, the data acquisition group GIP 
Ganil, the users support group and the accelerator staff of Ganil for 
their continuous support and efficient performances. They are also 
indebted to the technicians of  Smoluchowski Insitute of Physics and 
LPC Caen for their help during the construction and operation of the 
detector, and to the target laboratory of IReS Strasbourg and the 
isotope separator of CSNSM Orsay for the targets 
used in the experiments. 
This work was partly supported under the Polish-French IN2P3(CNRS)-KBN
agreement contract no. 101-01.


\begin{thebibliography}{10}

\bibitem{Hofm:95:1}
S.~Hofmann, V.~Ninov, F.~He$\beta$berger, P.~Armbruster, H.~Fogler,
  G.~M\"{u}nzenberg, H.-J. Sch\"{o}tt, A.~Popeko, A.~Yeremin, A.~Andreev,
  S.~Saro, R.~Janik, M.~Leino, Z. Phys. {\bf A350}, 277  (1995).

\bibitem{Oga:99:1}
Y.~Oganessian et~al., Eur. Phys. J.  {\bf A5}, 63 (1999).

\bibitem{Pete:01:1}
J.~P{\'e}ter et~al., in: G.~Fazio (Ed.), "Nuclear Physics et Border Lines,
  World Scientific, Lipari (Italy), 2001, pp. 257--266.

\bibitem{Zlo:00:1}
V.~Zlokazov, Eur. Phys. J.  {\bf A8}, 81 (2000).

\bibitem{Sch:00:1}
K.~H. Schmidt, Eur. Phys. J.  {\bf A8}, 141 (2000).

\bibitem{Zieg:85:1}
J.~F. Ziegler, J.~P. Biersack, U.~Littmark, The Stooping And Range Of Ions In
  Solids, Vol.~1, Pergamon Press, 1985.

\bibitem{Hube:89:1}
F.~Hubert et~al., Nucl. Ins. Meth.  {\bf B36}, 357 (1989).

\bibitem{england}
J.B.A.~England, Techniques in Nuclear Structure Physics, Mac Millan Press (Ed.), 1974.

\bibitem{Sosi:94:1}
Z.~Sosin, T.~Kozik, Z.~Majka, Nucl. Instr. Meth.  {\bf A351}, 383 (1994).

\bibitem{Loje:03:1}
K.~{\L}ojek, Report from Institute of Physics, Jagellonian University: 
in preperation.

\bibitem{Nomu:73:1}
T.~Nomura, K.~Hiruta, T.~Inamura, M.~Odera, Nuc. Phys.  {\bf A217},  253 (1973).

\end{thebibliography}

\centerline{\bf Figure captions.}
\par
\begin{figure}[htb]
\begin{center}
\includegraphics[scale=.8,angle=-90]{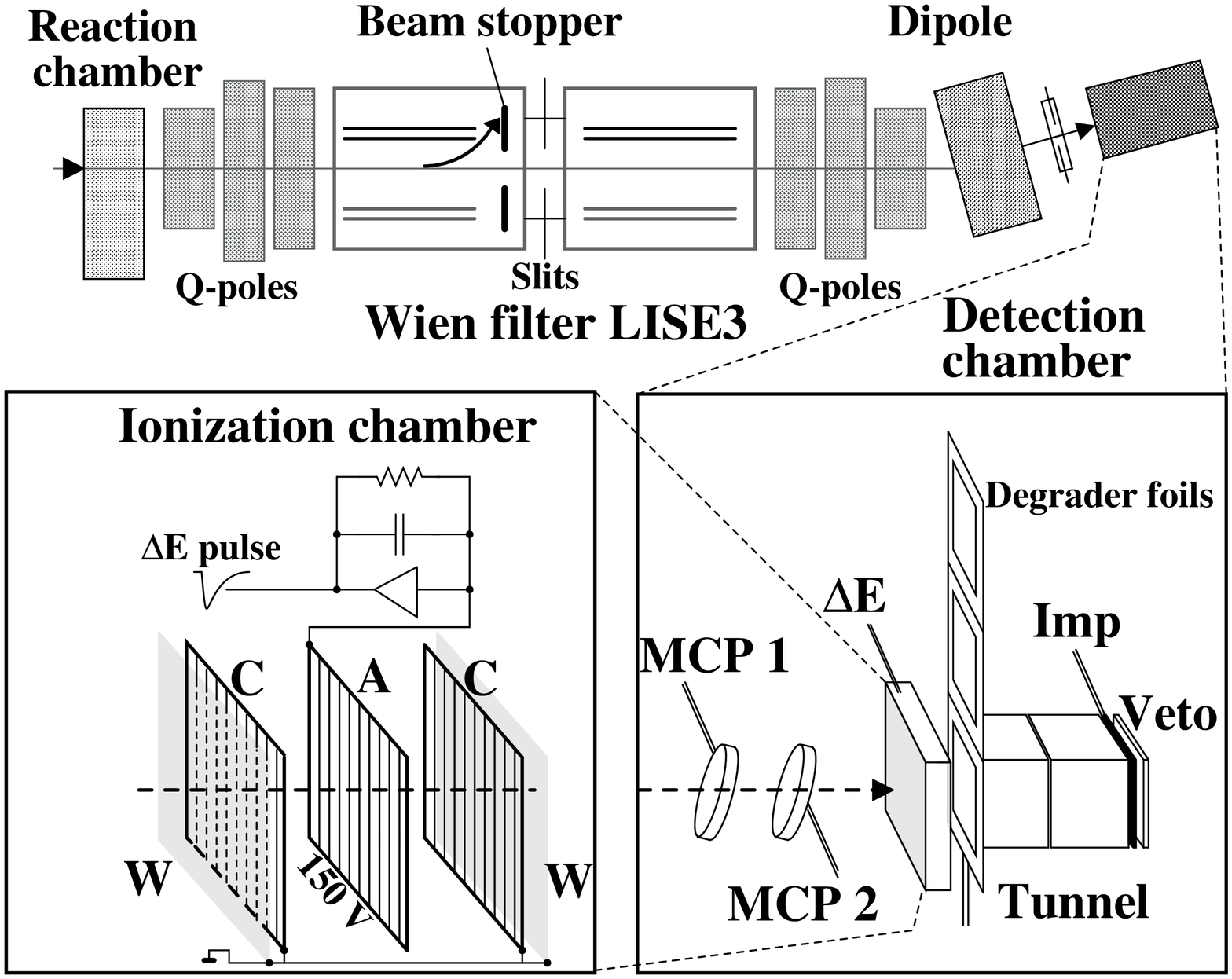}
\caption{}
\label{fig1}
\end{center}
\end{figure}

Fig.\ref{fig1} Schematic view of the experimental setup. Reaction chamber, two 
halves of the Wien filter as well as detection set-up is depicted here. Details of  
detection chamber are presented on the right panel. The left panel sketches  
more detailed construction of the ionization chamber: three wires planes 
cathode-anode-cathode and schematic electronics.
\vspace{0.5cm}
\par
\begin{figure}[htb]
\vspace{2cm}
\begin{center}
\includegraphics[scale=.8]{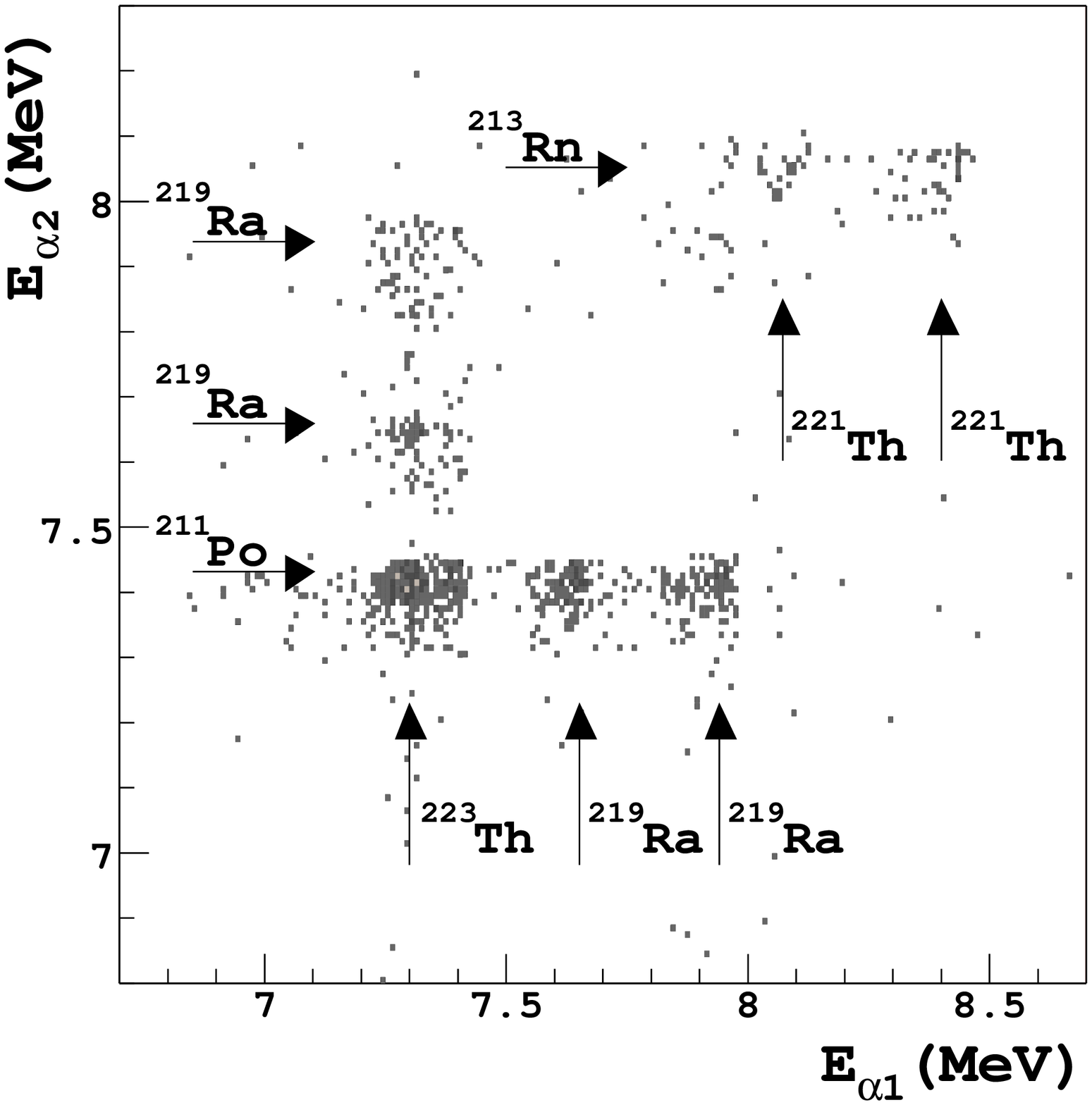}
\caption{}
\label{fig2}
\end{center}
\end{figure}

Fig.\ref{fig2} Energies of two detected $\alpha$ particles from the radioactive 
decay chains. The chains  are identified from the condition that $\alpha$'s are 
emitted from the same position as an  ER implanted in the strip detector (IMP). 
Observed groups of events (chains)  come from two ERs: $^{221}$Th and 
$^{223}$Th. The arrows in figure denotes $\alpha$'s which come from 
parent $^{223}$Th or $^{221}$Th, daughter $^{219}$Ra, grand-daughter 
$^{213}$Rn,  and grand-grand daughter $^{211}$Po nuclei. 
$^{208}$Pb+$^{18}$O reaction.
\vspace{0.5cm}
\par
\begin{figure}[htb]
\vspace{4cm}
\begin{center}
\includegraphics[scale=.8]{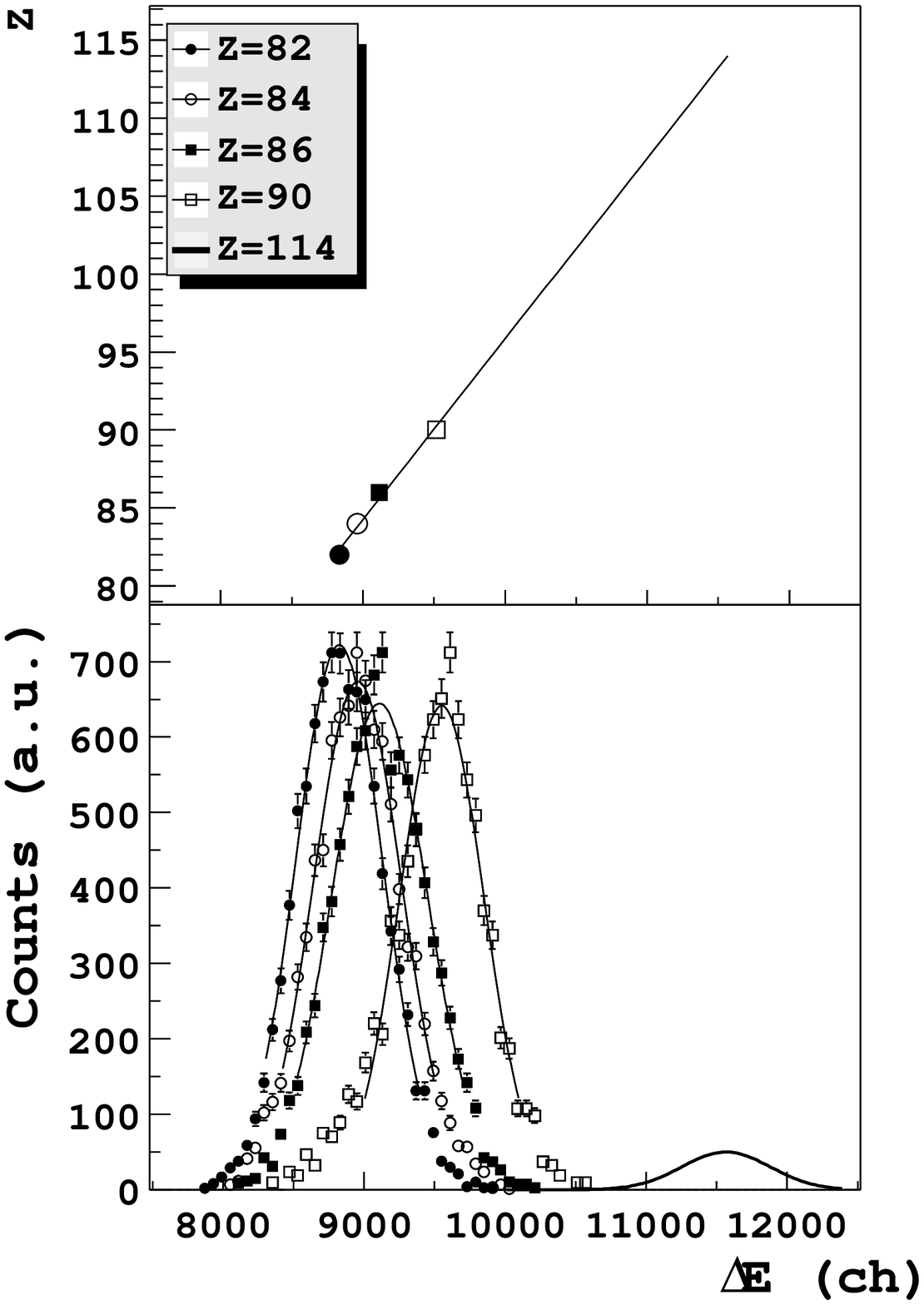}
\caption{}
\label{fig3}
\end{center}
\end{figure}

Fig.\ref{fig3} Lower panel: measured distributions of $\Delta$E signals for 
ions with different Z's. Full circles present energy loss distribution for Z=82 
while empty squares are for Z=90. Distribution denoted by empty circles and 
full squares are attributed (see text) to Z=84 and Z=86 respectively. The 
spectra were fitted with a gaussian to get the mean value and FWHM of 
$\Delta$E for each distribution.  In the upper panel the extracted mean values 
of $\Delta$E for each identified Z are presented. ``Calibration'' line shown in 
this panel is drawn through two points Z=82 and Z=90. It extrapolates to higher 
values of atomic number. The curve marked as Z=114 (lower panel) presents 
the predicted location of $\Delta$E signals for this super-heavy element.  
\vspace{0.5cm}
\par
\begin{figure}[htb]\vspace{2cm}
\begin{center}
\includegraphics[scale=.8]{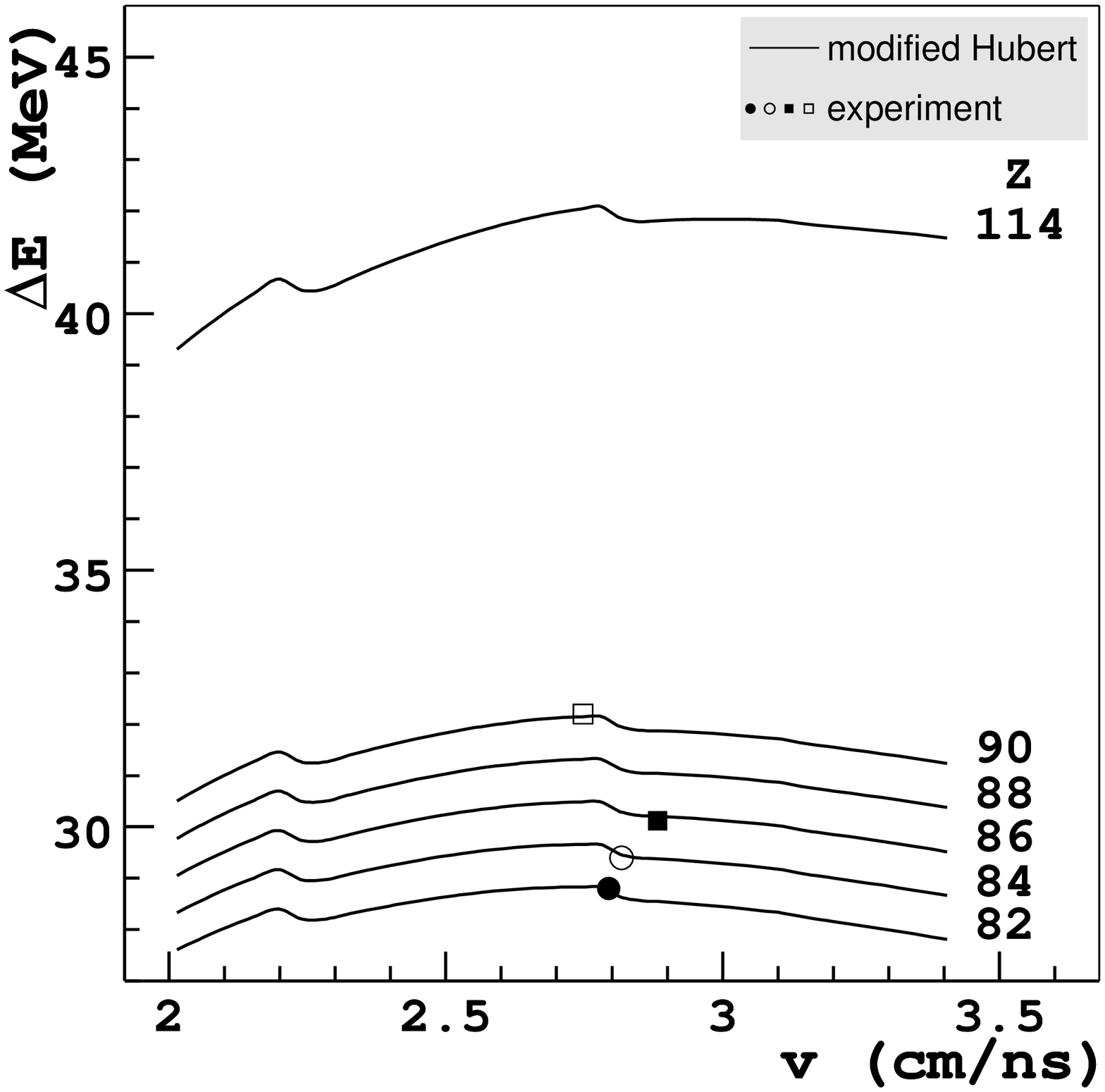}
\caption{}
\label{fig4}
\end{center}
\end{figure}

Fig.\ref{fig4} Energy loss in the ionization chamber for ions with atomic 
numbers marked in the figure as a function of their velocity, full lines.  
Calculation were done with a modified version of Hubert code. Symbols in the 
figures are experimental values of mean $\Delta$E for ions from 
fig.\ref{fig3}. They were normalized in the way that measured mean 
$\Delta$E for Z=82 corresponds to the calculated one.

\pagebreak

\pagebreak

\pagebreak
	
\pagebreak

\end{document}